\documentclass[aps,prl,twocolumn,showpacs]{revtex4}

\usepackage{graphicx}
\usepackage{amssymb}
\usepackage{epsfig}
\usepackage{subfigure}
\usepackage{amsmath}

\begin{document}

%

\title{\bf Large coupled oscillator systems 
with heterogeneous interaction delays}

\author{Wai Shing Lee, Edward Ott, Thomas M. Antonsen}

\affiliation{University of Maryland, College Park, MD 20742, U.S.A.}


\begin{abstract}
In order to discover generic effects of heterogeneous communication
delays on the dynamics of large systems of coupled oscillators,
this paper studies a modification of the Kuramoto
model incorporating a distribution of 
interaction delays. 
By focusing attention on the reduced
dynamics on an invariant manifold
of the original system, we derive governing
equations for the system which we use to study
stability of the
incoherent states and the dynamical transitional behavior from
stable incoherent states to stable coherent states.
We find that spread in the distribution function of delays can greatly 
alter the system dynamics.
\end{abstract}


\maketitle
Large systems of 
coupled oscillators occur in  
many situations in 
modern science and engineering \cite{PRK}. 
Noted examples include synchronous flashing of fireflies \cite{Buck}, 
pedestrian induced
oscillations of the Millennium Bridge \cite{MillB},
cardiac pace-maker cells \cite{Glass}, 
alpha rhythms in the brain \cite{Brain}, 
glycolytic oscillations in yeast populations \cite{Yeast}, 
cellular clocks governing circadian rhythm in mammals \cite{Yamaguchi},
oscillatory chemical reactions \cite{Kiss}, etc.
An incompletely
understood aspect
of such systems is that
signal propagation may take non-negligible
time, and that systems often have a finite reaction time to inputs
that they receive. 
Time delays
are thus both natural and inevitable in many of
these systems. 
In order to elucidate phenomena induced by time delay
in large coupled oscillator systems, Refs.\cite{YS,CKK} and \cite{Park}
carried out studies of globally coupled 
oscillators of the Kuramoto type \cite{Kuramoto} in the presence
of time delay.
These previous works all treated the special case in which
all time delays between interacting
oscillators were identical, and, in that context, they uncovered
many interesting behaviors revealing that time delay can 
profoundly affect the dynamics of coupled oscillator systems. 
However, in most situations where delays are an important
consideration,
the delays are not all identical.
The aim of this paper is to
study the more realistic case where there is a distribution of 
time delays along the links connecting the oscillators.
We shall see that previous striking features obtained in the case of uniform 
time delay are evidently strongly dependent on coherent communication between 
oscillators, and, as a consequence, are substantially changed by the 
incorporation of even modest spread in the time delays. 
For example, comparing results for typical cases with uniform delay and 
with a $30\%$ spread in delay, we will show that 
this delay spread 
(a) can completely eliminate the resonant structure 
in the average delay time dependence of the critical 
coupling $k_c$ for the onset of coherence, 
(b) can introduce hysteresis into the system behavior, 
and (c) can substantially 
decrease the number of attractors that simultaneously exist in a 
given situation.



We consider a network of oscillators with all-to-all coupling 
according to the classical Kuramoto scheme, but incorporating
link-dependent
interaction time delays $\tau_{ij}$
for coupling between any two oscillators $i$ and $j$,

\begin{equation}
\frac{d}{dt}\theta_i(t) = \omega_i + 
(k/N) 
\sum_{j=1}^{N} \sin
\left[ \theta_j(t - \tau_{ij}) - \theta_i(t)  \right],
\label{eq:Kura_dysp}
\end{equation}

\noindent
where $\theta_i(t)$ is the phase of oscillator $i$,
$\omega_i$ is the natural frequency of oscillator $i$, $k$ 
characterizes the coupling
strength between oscillators, $N$ is the total number of oscillators,
$\tau_{ii}=0$,
and $i=1,2 \cdots N$. 
Following Kuramoto, we note that the effect
of all the oscillators in the network on
oscillator $i$ may be expressed in terms of an ``order parameter''
$r_i$,
\begin{equation}
N^{-1} \sum_{j=1}^{N} \sin( \theta_j(t-\tau_{ij}) - \theta_i(t))
= \textrm{Im} [r_i e^{-i \theta_i(t)}],
\label{eq:Kura_desp_sup}
\end{equation}

\begin{equation}
r_i(t)= N^{-1} \sum_{j=1}^N e^{i \theta_j(t-\tau_{ij})}.
\label{eq:Kura_ord}
\end{equation}

\noindent
To facilitate the analysis,
we consider the following
two simplifying assumptions.
First, we consider the continuum limit $N \rightarrow \infty$
appropriate to the study of large systems, $N \gg 1$. 
Second,
we assume the collection of all
delays $\tau_{ij}$ is characterized by
a distribution $h(\tau)$
such that the fraction of links with delays between $\tau$
and $\tau + d\tau$ is $h(\tau) d\tau$. We, furthermore,
assume that, for randomly chosen links, $\tau$ is uncorrelated
with the oscillator frequencies $\omega$
at either end of the link.
These assumptions enable a description of the system dynamics 
in terms of a single oscillator distribution
function $f(\theta,\omega,t)$, which evolves in response 
to a mean field $r(t)$
according to the following
oscillator continuity equation,
\begin{equation}
\frac{\partial}{\partial t} f
+ \frac{\partial}{\partial \theta} 
\left\{  \left[ \omega + \frac{k}{2 i} 
        \left( e^{-i \theta}r - e^{i \theta} r^* \right)
           \right] f \right\} = 0.
\label{eq:Kura_FP}
\end{equation}

\noindent
In this case, the mean field $r(t)$ is given by
\begin{equation}
r(t) = \int_0^{\infty}  \xi(t-\tau) h(\tau) d\tau,
\label{eq:Kura_ord2}
\end{equation}

\begin{equation}
\xi(t) = \int_{-\infty}^{\infty} \int_{0}^{2 \pi}
   f(\omega,\theta,t) e^{i\theta} d\theta d\omega, 
\label{eq:Kura_xi}
\end{equation}

\noindent
where Eq.(\ref{eq:Kura_xi}) gives the input that nodes would receive 
in the absence of delay, and Eq.(\ref{eq:Kura_ord2}) ``corrects''
this input by incorporating the appropriate delay for each
fraction of inputing links, $h(\tau) d\tau$, with delay $\tau$.

Expanding $f(\omega,\theta,t)$ in a Fourier series, we have
\begin{displaymath}
f(\omega,\theta,t) = \frac{g(\omega)}{2 \pi}
\left\{ 1 +  \sum_{n=1}^{\infty} 
   \left[ f_n(\omega,t) e^{i n \theta} + f_n^*(\omega,t) e^{-i n \theta}
         \right] \right\}
\end{displaymath}

\noindent
where $g(\omega)\equiv \int_{0}^{2 \pi} f(\omega,\theta,t) d\theta$ is
the time-independent oscillator frequency distribution.
Following the method outlined in \cite{OA}, we consider the
dynamics of Eq.(\ref{eq:Kura_FP}) on an invariant manifold
in $f$-space:
\begin{equation}
f_n(\omega,t) = [a(\omega,t)]^n.
\label{eq:Lowd_def}
\end{equation}

\noindent
The macroscopic dynamics of $a(\omega,t)$ can
be derived by substituting Eq.(\ref{eq:Lowd_def})
into
Eq.(\ref{eq:Kura_FP}), leading to
\begin{equation}
\partial a/\partial t + i \omega a + (k/2)
\left( r a^2 - r^* \right) = 0.
\label{eq:Lowd_dyn}
\end{equation}

\noindent
In the case when the oscillator frequency distribution 
$g(\omega)$ is Lorentzian, i.e.,
\begin{equation}
g(\omega)=\frac{\Delta/\pi}{(\omega-\omega_0)^2 + \Delta^2},
\label{eq:Lorentzian}
\end{equation}

\noindent
and assuming suitable properties of the analytic continuation
into complex $\omega$ of $a(\omega,t)$ (see Ref.\cite{OA}),
Eq.(\ref{eq:Kura_xi}) can be evaluated explicitly by contour integration
with the contour closing at infinity
in the lower half complex $\omega$-plane to
give $\xi(t) = \int_{-\infty}^{\infty} g(\omega) a^*(\omega,t) d \omega
= a^*(\omega_0-i \Delta,t)$. Thus
Eq.(\ref{eq:Kura_ord2}) becomes
\begin{equation}
r(t)=\int_{0}^{\infty} a^*(t-\tau) h(\tau) d\tau.
\label{eq:Kura_ord2_Lor}
\end{equation}

\noindent
Furthermore, by setting
$\omega = \omega_0 - i \Delta$ in Eq.(\ref{eq:Lowd_dyn})
we have
\begin{equation}
\frac{d}{dt}a(t) + (\Delta + i \omega_0) a(t) + \frac{k}{2}
\left( r(t) a(t)^2 - r^*(t) \right) = 0,
\label{eq:Lowd_dyn_Lor}
\end{equation} 

\noindent
where in both Eqs.(\ref{eq:Kura_ord2_Lor}) and (\ref{eq:Lowd_dyn_Lor})
the particular argument value $\omega = \omega_0 - i \Delta$
has been suppressed; i.e., $a(\omega_0 - i \Delta,t)$ is replaced
by $a(t)$. 
Equations (\ref{eq:Kura_ord2_Lor}) and (\ref{eq:Lowd_dyn_Lor}) 
thus form a complete description for the 
dynamics on the invariant manifold (\ref{eq:Lowd_def})
when $g(\omega)$ is Lorentzian.
Recently a result has been obtained \cite{OA_2} that, when applied
to our problem, establishes that all attractors of the full system,
Eqs.(\ref{eq:Kura_FP}) - (\ref{eq:Kura_xi}), are also 
attractors of our reduced system,
Eqs.(\ref{eq:Kura_ord2_Lor}) and (\ref{eq:Lowd_dyn_Lor}), and vice versa.
(The result of Ref.\cite{OA_2} was previously strongly indicated by numerical
experiments of Ref.\cite{Martens}.)


Previous studies of the effect of delay on the Kuramoto system 
(Refs.\cite{YS,CKK,Park}) considered uniform delay on all the links,
corresponding to $h(\tau)=\delta(\tau-T)$.
Our goal is to uncover the effect of heterogeneity of delays
along the network links. Accordingly, we consider that
$h(\tau)$ has some average value $T$ with a spread about
this value, and $h(\tau) \equiv 0$ for $\tau < 0$. A convenient
class of functions for this purpose is 
\begin{equation}
h(\tau) = \frac{1}{T} \hat{h}_n \left( \frac{\tau}{T} \right),
\mbox{where} \hspace{1mm} 
\hat{h}_n(\hat{\tau}) = A_n \hat{\tau}^n e^{-\beta_n \hat{\tau}}.
\label{eq:delay_distF}
\end{equation}

\noindent
Here, $A_n$ and $\beta_n$ are determined by the normalization 
conditions: $\int_0^{\infty} \hat{h}_n(\hat{\tau}) d\hat{\tau}=1$ and 
$\int_0^{\infty} \hat{\tau }\hat{h}_n(\hat{\tau}) d\hat{\tau}=1$,
yielding
\begin{equation}
A_n = (n+1)^{n+1}/n! \hspace{2mm} ,\hspace{2mm}
\beta_n = n+1.
\label{eq:delay_dist_para}
\end{equation}

\noindent
For this family of distributions, we have that the standard deviation
of $\tau$ about its mean $T$ is given by
\begin{equation}
\delta \tau = (<\tau^2>-<\tau>^2)^{1/2}  
            = T / \sqrt{n+1}.
\label{eq:delay_dist_var}
\end{equation}

\begin{figure}[h]
  \begin{center}
    \epsfig{file=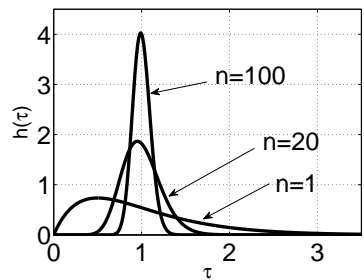,width=150pt}
    \caption{Graphs of $h(\tau)$ at $T=1$ for $n=1,20,100$.}
    \label{fig:hn_tau1_vary_n}
  \end{center}
\end{figure}

\noindent
Thus, for $n \rightarrow \infty$, we recover the 
case, $h(\tau)=\delta(\tau-T)$,
previously investigated in Refs.\cite{YS,CKK,Park}.
And, by decreasing $n$, we can study the effect of increasing the
relative spread $\delta \tau / T$ in the delay times. The
dependence of $h(\tau)$ on $n$ is depicted in Fig.\ref{fig:hn_tau1_vary_n}.

We can exploit the convolution
form of (\ref{eq:Kura_ord2}) 
and turn it into a differential equation
for $r(t)$. Taking a Laplace Transform,
we have, for the case of Lorentzian $g(\omega)$,
\begin{equation}
\bar{r}(s) 
= H(s) \bar{a}^*(s),
\label{eq:Lowd_r_freq}
\end{equation}

\noindent
where $\bar{r}(s)$ and $\bar{a}(s)$
are the Laplace transform of 
$r(t)$ and $a(t)$ respectively, while
\begin{equation}
H(s) = [(T/\beta_n)s+1]^{-(n+1)},
\label{eq:La_h}
\end{equation}

\noindent
is the Laplace transform of $h(\tau)$.
Our choice of the function class given by Eq.(\ref{eq:delay_distF})
is motivated by the fact that it yields a particularly 
convenient Laplace transform and corresponding time-domain formulation.
In particular, transforming back to the time-domain by
letting $s \rightarrow d/dt$, Eq.(\ref{eq:Lowd_r_freq}) yields
\begin{equation}
\left[ (T/\beta_n) (d/dt) + 1 \right]^{n+1} r(t) = 
a^*(t).
\label{eq:Lowd_r}
\end{equation}

\noindent
Thus, we now have Eqs.(\ref{eq:Lowd_dyn_Lor}) and 
(\ref{eq:Lowd_r}) as our
description for the dynamics on the invariant manifold
with heterogeneous link delays.
Here, it is noteworthy that Eqs.(\ref{eq:Lowd_dyn_Lor}) and
(\ref{eq:Lowd_r}) form a system of ordinary
differential equations
in comparison with the original system Eq.(\ref{eq:Kura_dysp}) which
comprises a very large number of 
time-delay differential equations. 
Note that for the case of uniform delay, 
$h(\tau)=\delta(\tau-T)$,
we take the limit
$n \rightarrow \infty$,
in which case
Eq.(\ref{eq:Lowd_r_freq}) takes the form 
$\bar{r}(s)=e^{-sT} \bar{a}^*(s)$, 
yielding $r(t) = a^*(t-T)$, which, when
substituted  
into Eq.(\ref{eq:Lowd_dyn_Lor}), gives
the time-delay differential equation for $a(t)$ in Ref.\cite{OA}.


A trivial exact solution to the system
(\ref{eq:Lowd_dyn_Lor}) and (\ref{eq:Lowd_r}) is given by
$r(t) = a(t) = 0$, which we refer to as the 
``incoherent state'' \cite{note1}.
Stability of the incoherent state
can be studied by linearizing Eq.(\ref{eq:Lowd_dyn_Lor}) about
the solution $a(t)=0$ and
setting $a(t)=a_0 e^{st}$, from which
we obtain
\begin{equation}
1 = [k H(s)/2]  (s+i \omega_0+\Delta)^{-1}.
\label{eq:Disp_incoh}
\end{equation}

\noindent
The critical coupling $k_c$ at
which a stable incoherent state
solution becomes unstable as $k$ increases through $k_c$,
corresponds a solution to Eq.(\ref{eq:Disp_incoh})
with $\textrm{Re}(s)=0$. 

\begin{figure}[h]
  \begin{center}
    \epsfig{file=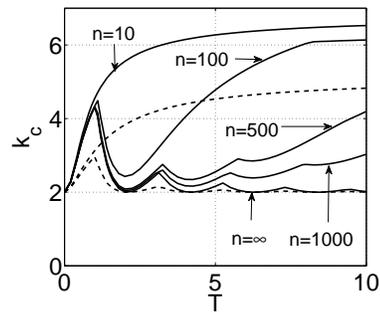,width=150pt}
    \caption{Solid curves are plots of the 
             critical value of $k_c$ versus $T$ for 
             Lorentzian $g(\omega)$ with $\omega_0=3,\Delta=1$, and
             $n=10,100,500,1000,\infty$, corresponding to 
             $\delta \tau/T \approx 0.302,0.1,0.045,0.032,0$. The dashed
             curves are for Gaussian $g(\omega)$ as described in the text.}
    \label{fig:incoh_sta_data}
  \end{center}
\end{figure}

The solid curves in
Fig.\ref{fig:incoh_sta_data}
show results obtained from
Eq.(\ref{eq:Disp_incoh}) with Lorentzian $g(\omega)$ for
the critical coupling value $k_c$ versus $T$ at different
$n$'s with parameters $\omega_0=3$ and $\Delta = 1$. 
For the case of uniform delays ($n \rightarrow \infty$),
$k_c$ as a function of $T$ exhibits
the type of dependence found in
Ref.\cite{YS}
with characteristic ``resonances''. However, as the relative
spread $\delta \tau / T$ is increased ($n$ is decreased), we
see that the resonant structure that applies for the case of
zero spread is strongly modified. For example, even 
at the relatively small spread of $\delta \tau / T \approx 0.1$
(corresponding to $n=100$), there is only one peak (at $T \approx 1$)
and one minimum (at $T \approx 2$), with $k_c$ for $T > 2$ being
very substantially higher than in the case of no spread.
For $\delta \tau / T \approx 0.302$ ($n=10$) the effect is 
even more severe, and the previous resonant structure is 
completely obliterated. For comparison, the dashed curves in
Fig.\ref{fig:incoh_sta_data} show results for 
$\delta \tau/T \approx 0.302$ (upper)
and $0$ (lower) when $g(\omega)$ is Gaussian with the same peak value 
as for the Lorentzian distribution used to obtain the solid
curves \cite{note3}. The Gaussian and Lorentzian results are similar, 
suggesting that the qualitative behavior does not depend strongly
on details of $g(\omega)$.

\begin{figure}[h]
  \begin{center}
    \subfigure []
    {\epsfig{file=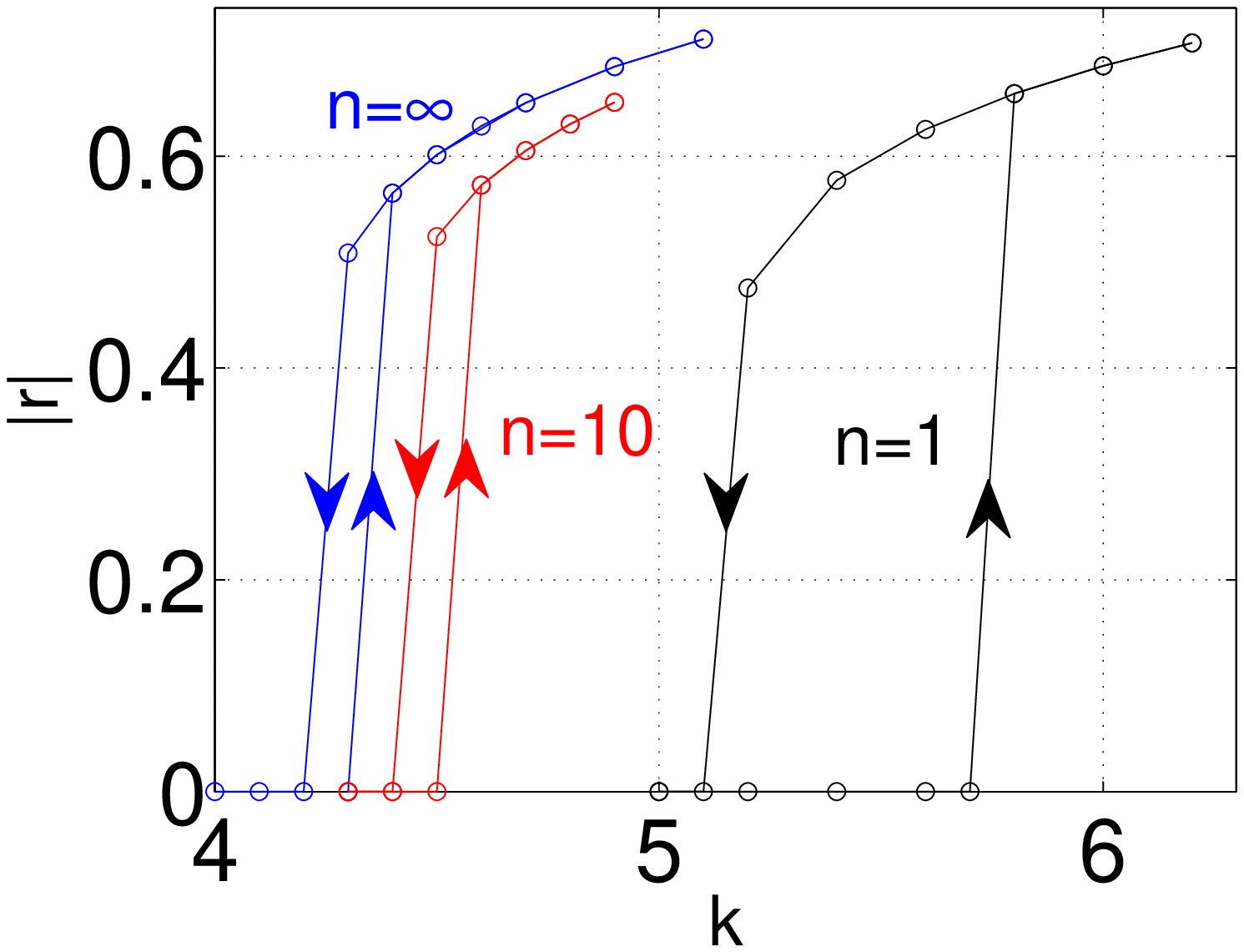,width=118pt} \label{fig:hyste_1}}
    \subfigure []
    {\epsfig{file=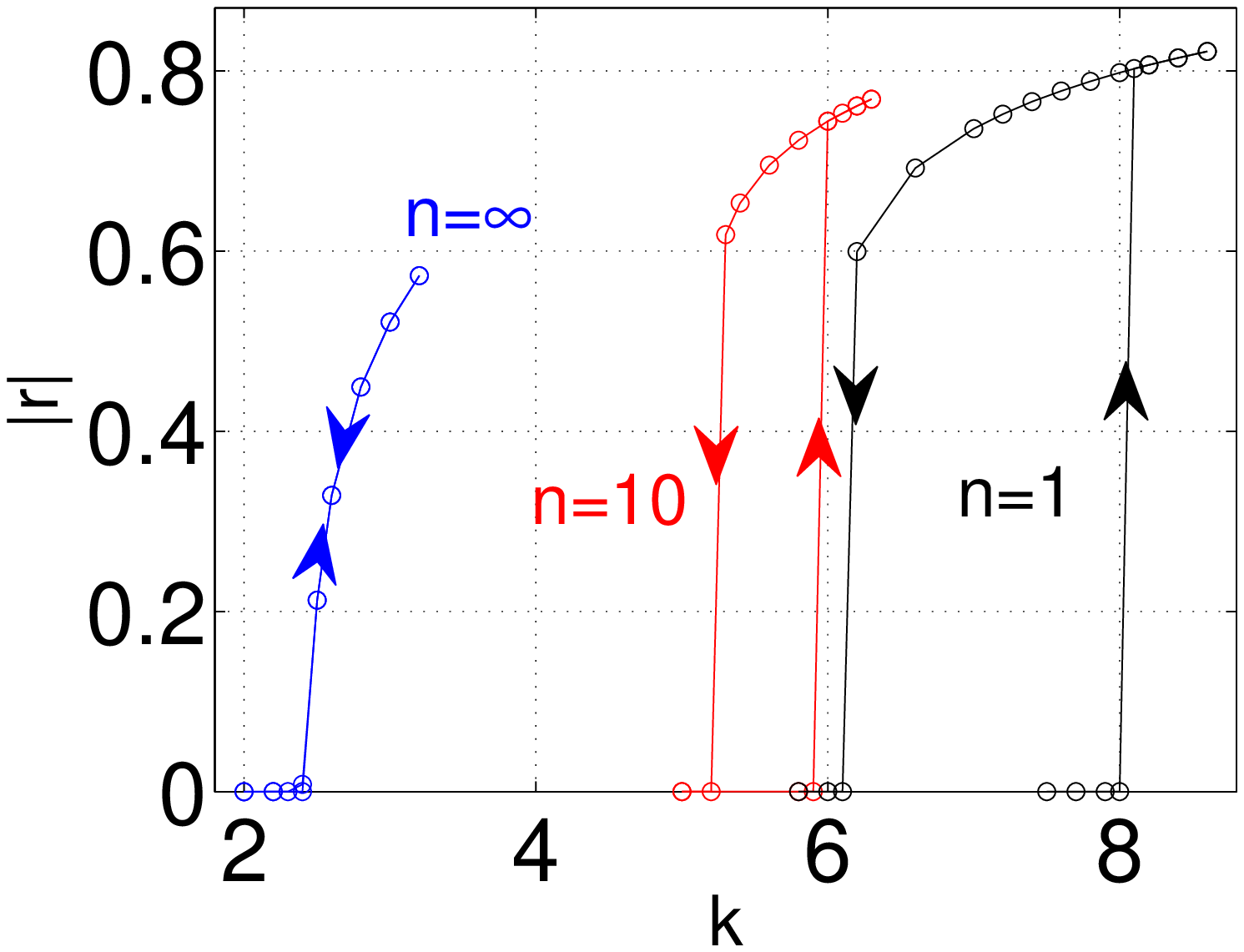,width=118pt} \label{fig:hyste_2}}
    \caption{(Color online) Hysteresis loop at $\omega_0=3$, $\Delta=1$, (a) for $T=1$,
             (b) for $T=3$.}
    \label{fig:hyste}
  \end{center}
\end{figure}

As reported in Ref.\cite{YS}, bistable behavior can exist;
i.e., a situation in which both incoherent
and coherent states are stable.
In Figs.\ref{fig:hyste_1} and \ref{fig:hyste_2} we show the
hysteresis loops obtained by numerical solution
of Eqs.(\ref{eq:Lowd_dyn_Lor}) and (\ref{eq:Lowd_r})
for $n < \infty$ and, for $n=\infty$, where 
the $n=\infty$ result is obtained by solution
of the delay equation obtained by inserting
$r(t)=a^*(t-T)$ in (\ref{eq:Lowd_dyn_Lor}).
Comparing Fig.\ref{fig:hyste_1}, which is for $T=1$, with 
Fig.\ref{fig:hyste_2}, which is for $T=3$, we note the striking 
result that,
for large $T$, hysteresis is sustained only
with large enough spread in the delay distribution, i.e., 
when $n$ is small [e.g., for $n=\infty$ and $T=3$ (Fig.\ref{fig:hyste_2}) the
bifurcation is supercritical and hysteresis is absent].

Coherent oscillatory states can be obtained by substituting
the ansatz $a(t)=a_0 e^{i \Omega t}$, where
$a_0$ and $\Omega$ are real constants,
into Eq.(\ref{eq:Kura_ord2_Lor}) 
and (\ref{eq:Lowd_dyn_Lor}). This gives
\begin{align}
\left[ i(\Omega + \omega_0) + \Delta \right] +
& (k/2) \left( a_0^2 H^*(i \Omega) - H(i \Omega) \right) = 0, 
\nonumber \\
r(t) &= a_0 e^{-i \Omega t} H^*(i \Omega).
\label{eq:Coh}
\end{align}

\begin{figure}[htb]
  \begin{center}
    \subfigure[]
    {\epsfig{file=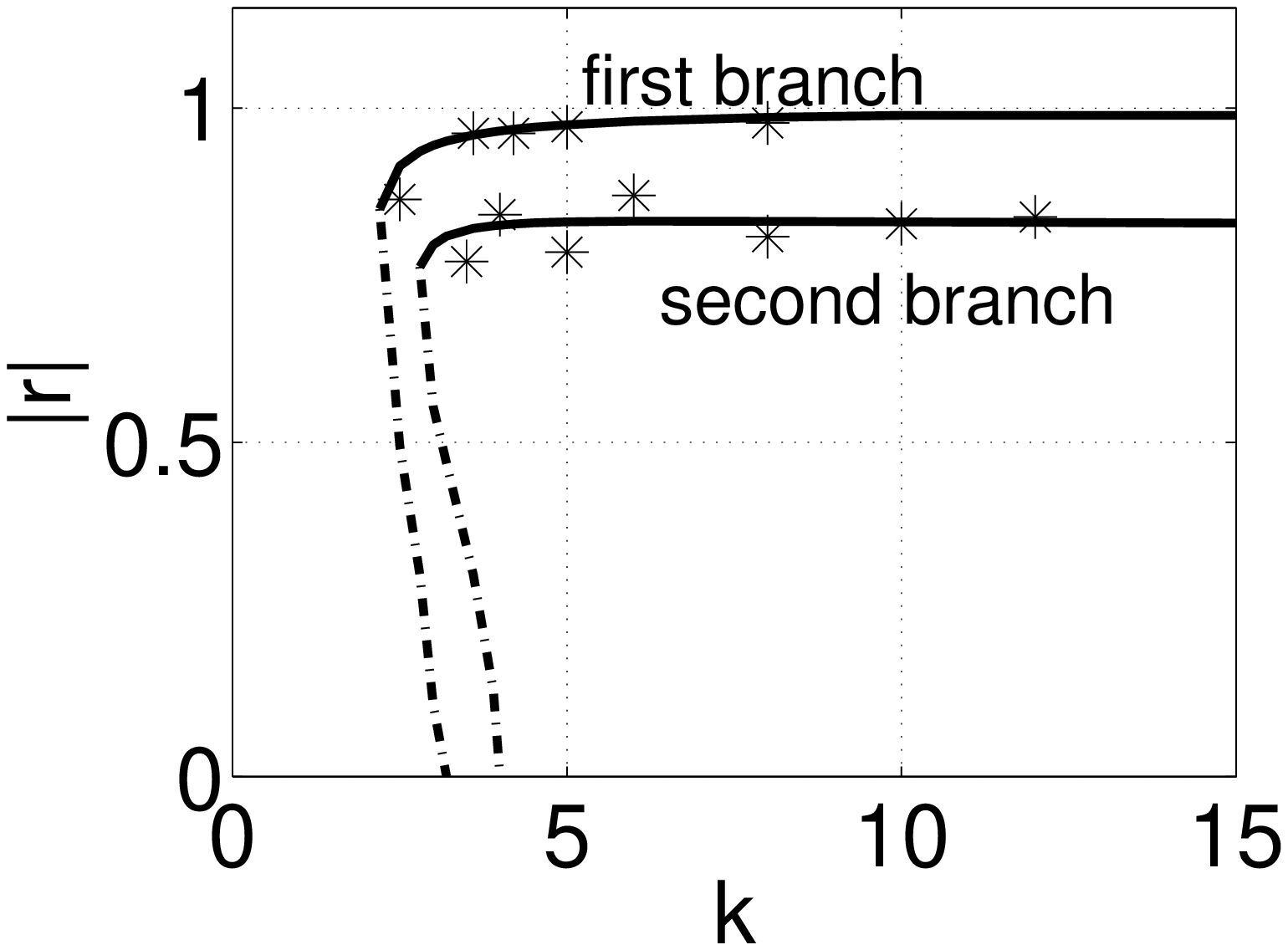,width=118pt} \label{fig:coh_r}}
    \subfigure[]
    {\epsfig{file=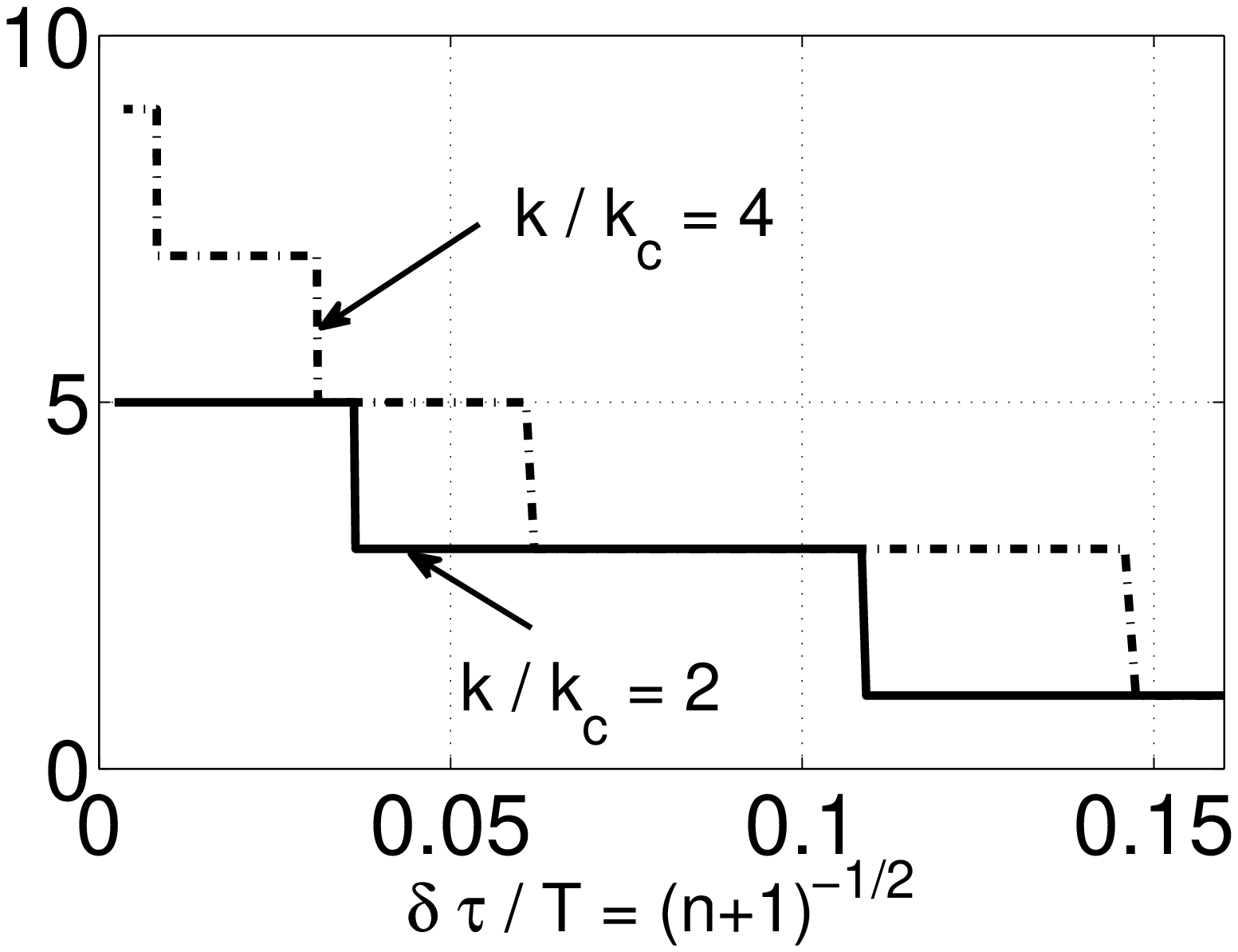,width=118pt} \label{fig:num_coh}} 
    \caption{(a) Magnitude of $r$ for the first two branches of 
             coherent states with parameter values:
             $\omega_0=3,\Delta=0.1$, $n=100,T=1$ for $h(\tau)$. 
             Solid lines give the theoretical
             values of the stable coherent states, dashed lines give the 
             unstable coherent 
             states, and asterisks give the simulation results.
             (b) Number of coherent attractors
             (number of solutions of Eq.(\ref{eq:Coh}) that are stable 
             according to (\ref{eq:Coh_stab}))
             versus $\delta \tau / T$
             for the following parameters:
             $k=40, T=1, \omega_0=0$; $\Delta=10$ for the solid line 
             ($k_c=20$),
             $\Delta=5$ for the dashed line ($k_c=10$).}
    \label{fig:coh_state}
  \end{center}
\end{figure}

\noindent
As reported in both
\cite{YS} and \cite{CKK}, for $h(\tau) = \delta(\tau-T)$, 
multiple branches of coherent state solutions
are possible in Eq.(\ref{eq:Coh}). Furthermore, we can employ
Eqs.(\ref{eq:Lowd_dyn_Lor}) and (\ref{eq:Lowd_r})
to study
the stability of each coherent state 
by introducing a small perturbation $\delta a(t) e^{i \Omega t}$
to the coherent state solution in (\ref{eq:Coh}), with
$\delta a(t) = K_1 e^{s t} + K_2 e^{s^* t}$.
This yields the 
following equation for $s$:
\begin{equation}
\begin{array}{l}
[ s - \frac{k}{2}H(s+i\Omega) + A ][ s - \frac{k}{2}H(s-i\Omega) + A^* ] \\
= \left( k a_0^2 / 2 \right)^2 H(s-i\Omega)H(s+i\Omega),
\end{array}
\label{eq:Coh_stab}
\end{equation}

\noindent
where $A=\Delta + i(\omega_0 + \Omega) + k a_0^2 H(-i \Omega)$.
Instability of each coherent state is then determined by
whether there are solutions to (\ref{eq:Coh_stab}) for $s$ with
positive real parts. In Fig.\ref{fig:coh_r}
we compare the theoretical results 
for $|r|$ calculated from
Eqs.(\ref{eq:Coh}) and (\ref{eq:Coh_stab})
with simulation results based on Eq.(\ref{eq:Kura_dysp}) with
$N=100$ and $\delta \tau /T \approx 0.1$ ($n=100$) for
the first two branches of coherent states
with $\omega_0=3,\Delta=0.1,T=1$.
The solid (dashed) curves correspond to stable (unstable) coherent states. 
The Eq.(\ref{eq:Kura_dysp}) simulation values reported
in the figures represent time averages of these quantities computed
after the solution has apparently settled into the coherent state.
It is seen that there is good agreement between the theory and
simulations using Eq.(\ref{eq:Kura_dysp}).
In addition, on simulating these two branches 
of coherent states, we
verified that the finding of Ref.\cite{CKK}
that the basin of attraction is large for the first branch, but small
for the second one, also holds with heterogeneous delays.

Furthermore,
the number of coherent attractors
strongly depends on the spread in delay times.
Figure \ref{fig:num_coh} shows
the dependence of
the number of coherent attractors
on the relative delay spread 
$\delta \tau / T = (n+1)^{-1/2}$,
with $k=40, \omega_0=0, T=1$, for two values of the frequency spread,
$\Delta=5$ (dashed) and $\Delta=10$ (solid) (for which
$k_c=10$ and $20$, respectively).
For both cases, it is seen that as the relative delay spread is
increased ($(n+1)^{-1/2}$ is increased), 
the number of coherent attractors decreases. And there 
remains at least one such attractor when $n$ approaches unity, while
a parameter dependent maximum is attained
when $n \rightarrow \infty$, which we find is generally larger for 
smaller $\Delta$ and larger $k$ \cite{note2}.

In conclusion,
in this paper we address, for the first time, the effect of
heterogeneous delays on the dynamics of globally coupled phase
oscillators. As compared to the case
of uniform delay (Refs.\cite{YS,CKK,Park}), we find that
delay heterogeneity can have important consequences, among
which are the following:
(i) decrease in resonant structure of the dependence of $k_c$ on $T$ 
(Fig.\ref{fig:incoh_sta_data});
(ii) increase of $k_c$ 
(Fig.\ref{fig:incoh_sta_data});
(iii) enhancement of hysteretic effects 
(Figs.\ref{fig:hyste_1} and \ref{fig:hyste_2});
(iv) reduction in the number of coherent attractors (Fig.\ref{fig:num_coh}).
Furthermore, we have introduced a framework for the study
of delay heterogeneity that can be readily applied to a
variety of extensions of the Kuramoto model, such as
communities of oscillator populations with different
community dependent characteristics \cite{Barretto},
non-monotonic $g(\omega)$ \cite{Martens}, and periodic driving
\cite{Antonsen} (see Ref.\cite{OA} for more examples).

This work was supported by ONR award N00014-07-0734 and by 
NSF (Physics).



\begin{thebibliography}{99}
\bibitem{note1}
Note that for $r(t)=0$, Eq.(\ref{eq:Kura_FP}) implies that the oscillators
do not interact, and run freely at their natural frequencies whose
spread leads to a uniform distribution in phase.
\end{thebibliography}
\end{document}